\documentclass{acm_proc_article-sp}

\usepackage{graphicx}
\usepackage{multirow}
\usepackage[table]{xcolor}
\usepackage{array}
\usepackage{epstopdf}
\usepackage{listings}
\usepackage{amsmath}
\usepackage[htt]{hyphenat}
\usepackage{subfigure}
\usepackage{color}
\usepackage{pxfonts}

\begin{document}

%\lefthyphenmin4
%\righthyphenmin3
\definecolor{mygreen}{rgb}{0,0.6,0}
\lstdefinestyle{mystyle}{
breaklines=true,
basicstyle=\ttfamily,
frame=single,
captionpos=b,
commentstyle=\itshape\color{mygreen},
keywordstyle=\bfseries,
language=Java,
stepnumber=1,
numbers=left,
firstnumber=1,
numberfirstline=true,
numbersep=2pt,
framexleftmargin=1.5em,xleftmargin=2em,
numberstyle=\footnotesize
}

\lstset{style=mystyle} % Font style for Coded text.

\title{Teaching Parallel Programming Using Java}

%
% You need the command \numberofauthors to handle the 'placement
% and alignment' of the authors beneath the title.
%
% For aesthetic reasons, we recommend 'three authors at a time'
% i.e. three 'name/affiliation blocks' be placed beneath the title.
%
% NOTE: You are NOT restricted in how many 'rows' of
% "name/affiliations" may appear. We just ask that you restrict
% the number of 'columns' to three.
%
% Because of the available 'opening page real-estate'
% we ask you to refrain from putting more than six authors
% (two rows with three columns) beneath the article title.
% More than six makes the first-page appear very cluttered indeed.
%
% Use the \alignauthor commands to handle the names
% and affiliations for an 'aesthetic maximum' of six authors.
% Add names, affiliations, addresses for
% the seventh etc. author(s) as the argument for the
% \additionalauthors command.
% These 'additional authors' will be output/set for you
% without further effort on your part as the last section in
% the body of your article BEFORE References or any Appendices.

\numberofauthors{3} %  in this sample file, there are a *total*
% of EIGHT authors. SIX appear on the 'first-page' (for formatting
% reasons) and the remaining two appear in the \additionalauthors section.
%
\author{
% You can go ahead and credit any number of authors here,
% e.g. one 'row of three' or two rows (consisting of one row of three
% and a second row of one, two or three).
%
% The command \alignauthor (no curly braces needed) should
% precede each author name, affiliation/snail-mail address and
% e-mail address. Additionally, tag each line of
% affiliation/address with \affaddr, and tag the
% e-mail address with \email.
%
% 1st. author
\alignauthor Aamir Shafi\titlenote{aamir.shafi@seecs.edu.pk},
             Aleem Akhtar\titlenote{aleem.akhtar@seecs.edu.pk},
             Ansar Javed\titlenote{ansar.javed@seecs.edu.pk}\\
 \affaddr{SEECS, National University of Sciences and Technology (NUST), Pakistan}\\
 %\email{\{aamir.shafi,aleem.akhtar,ansar.javed\}@seecs.edu.pk}
% 2nd. author
\alignauthor Bryan Carpenter\titlenote{bryan.carpenter@port.ac.uk}\\
 \affaddr{School of Computing, University of Portsmouth (UK)}
 }
% There's nothing stopping you putting the seventh, eighth, etc.
% author on the opening page (as the 'third row') but we ask,
% for aesthetic reasons that you place these 'additional authors'
% in the \additional authors block, viz.

% Just remember to make sure that the TOTAL number of authors
% is the number that will appear on the first page PLUS the
% number that will appear in the \additionalauthors section.

\maketitle
\begin{abstract}
This paper presents an overview of the ``Applied Parallel Computing''
course taught to final year Software Engineering undergraduate students
in Spring 2014 at NUST, Pakistan. The main objective of the course was
to introduce practical parallel programming tools and techniques for
shared and distributed memory concurrent systems. A unique aspect of
the course was that Java was used as the principle programming language.
The course was divided into three sections. The first section covered
parallel programming techniques for shared memory
systems that include multicore and Symmetric Multi-Processor (SMP) systems.
In this section, Java threads was taught as a viable programming API for
such systems. The second section was dedicated to parallel programming
tools meant for distributed memory systems including clusters and
network of computers. We used MPJ Express---a Java MPI library---for
conducting programming assignments and lab work for this section.
The third and the final section covered advanced topics including
the MapReduce programming model using Hadoop and the General Purpose
Computing on Graphics Processing Units (GPGPU).

\end{abstract}

%A category including the fourth, optional field follows...


% NOT required for Proceedings

\section{Introduction}
\label{sec:Introduction}

%In the last decade, the software industry has witnessed a sea change
%where single power hungry cores are making way for multiple power
%efficient processing cores. The main reason is that increasing the clock
%speed exponentially increases power consumption and heat dissipation of
%the processor. As a consequence major microprocessor vendors like Intel,
%AMD, Sun, and IBM have shifted their business model to increasing cores
%instead of increasing clock speed \cite{herb}. Following this trend, many High
%Performance Computing (HPC) vendors are also building clusters using
%multicore processors now.

The emergence of multicore hardware has brought parallel computing into
the limelight. It has also put the burden of improving performance of
applications on the software programmers \cite{herb}. The only option to increase
performance of existing sequential applications is to utilize some form
of parallelism. This obviously implies that the software development
community---including current and future software engineers---must
learn parallel programming models and tools
to write optimized code for multicore processors and High Performance
Computing (HPC) hardware.

%Leveraging the raw performance of modern multicore processors and
%emerging  heterogeneous computing systems prevalent in the industry
%requires today's  software developers to understand the theoretical and
%practical underpinnings  of these modern systems. This evidently poses
%a challenge to universities to produce students that are well equipped
%with the concepts of parallel computing. To engage universities to develop
%and teach parallel computing courses and materials, the High Performance
%Computing (HPC) community at large has initiated activities to meet
%this goal. These activities include: 1) The education in HPC section in
%top Supercomputing Conferences like, SC, IPDPS and ISC. 2) The student
%cluster competitions to not only create interest for HPC in students
%but also provide hands on experience with modern HPC architectures. 3)
%The NSF/IEEE-TCPP Curriculum Initiative, Shodor and CSinParallel to act
%as a guideline and point of collaboration with in HPC educators.

Realizing the importance of teaching concurrency at the undergraduate
 level, a $2+1$ credit hours elective course titled ``Applied Parallel
 Computing'' was added to the Bachelors of Software Engineering
 program\footnote{http://seecs.nust.edu.pk/academics/doc/bese.php} at NUST, Pakistan. The program
spans four years---distributed in eight  semesters---and $136$ credit
 hours. This particular course on parallel  computing was taught in
the eighth and the last semester. Course contents were
 mostly adapted from a Parallel and Distributed Computing (PDC)
course  taught at the University of Portsmouth, UK.

%
% put reference to this link for Software Engineering program
% http://seecs.nust.edu.pk/academics/doc/bese.php
%

%Following this endeavour our universities---National University of
%Sciences and Technology in Pakistan and University of Portsmouth in
%UK---have been offering courses and workshops on parallel computing to
%students and industry, respectively.

The course began with an introduction of parallel computing, which
 motivated the need for such computing to solve some of the biggest
possible problems  in the least possible time. Some important concepts
including  shared/distributed  memory systems, performance measurement
 metrics, and hardware accelerators  were introduced. After the initial
 introduction, the course was  divided into three sections. The
 first section covered programming techniques for shared memory
systems including multicore processors and Symmetric Multi-Processors
(SMPs). These included Java threads \cite{threadsJava}, OpenMP \cite{openmp},  and Intel
Cilk Plus \cite{cilkplus}. Note that all practical work including assignments,
labs, and code samples during lectures were Java-based.
%The  reason for
%covering non-Java technologies like OpenMP/Intel Cilk Plus  in the theory
%part was to provide complete coverage of shared memory  programming
%technologies.
The second section covered programming tools and APIs
for distributed memory systems including commodity clusters. For
this section, the course focused on writing parallel applications
using a Java MPI-like software called MPJ Express \cite{Shafi2009},
which  implements the mpiJava $1.2$ API specification \cite{carpenter}---this is
equivalent  to MPI version $1.1$. Being a Java MPI library, MPJ Express
allows writing parallel Java applications for  clusters and network of
computers. To allow easy configuration and  installation for students,
MPJ Express provides a {\em multicore mode}, where  Java threads are
used to simulate parallel MPI processes in a single JVM. Our students
found this to be an extremely useful feature since initially they were
able to write, execute, and test parallel Java code on their personal
laptops/PCs. Once stable, the same code would also execute on a cluster
or a network of computers using the MPJ Express {\em cluster mode}. In
the second section on distributed memory systems programming, a range  of
topics including synchronous/asynchronous point-to-point communication,
collective communication, and synchronization were covered. The third
and the  final section covered advanced topics including the MapReduce
programming model using Hadoop and the General Purpose Computing on
Graphics Processing Units (GPGPU).

% see if we can use this
%{\color{red} The final lectures were delivered by experts from local HPC
%industry which introduced students to the kind of HPC work being done
%locally. It also gave them a chance to interact with future potential
%employers.}

\subsection{Motivation for using Java}

% see if we can use this
%{\color{red} High level languages have become choice of teaching computer programming, algorithms and data structures to students.}

An interesting and unique aspect of this parallel computing course was
preferring Java over traditional native HPC languages like C and Fortran
for the practical part of the course. There are several reasons for this.

%Almost immediately following its release in $1996$ Java became a
%``mainstream'' programming language of the software industry.
Compared with C or Fortran, the advantages of the Java programming
language include higher-level programming concepts, improved compile
time and runtime checking, and, as a result, faster problem detection
and debugging. In addition, Java's automatic garbage collection,
when exploited carefully, relieves the programmer of many of the
pitfalls of lower-level languages. The built-in support for threads
provides a way to insert parallelism in Java applications.
The Java Development Kit (JDK) includes a large set of libraries
that can be reused by developers for rapid application
development. Another, interesting, argument in favour of Java is
the large pool of developers---the main reason is that Java is
taught as one of the major languages in many Universities around
the globe. A highly attractive feature of applications written in Java is that
they are portable to any hardware or operating system, provided
that there is a Java Virtual Machine (JVM) for that system. The
contribution of the JVM is significant, keeping in mind that it
allows programmers to focus on issues related to their application
and domain of interest, and not on system heterogeneity.

In order to facilitate writing parallel Java code for shared memory
systems, Java is equipped with a feature-rich
threading API. In order to teach programming distributed memory systems,
we had the option to choose between various Java MPI libraries including
MPJ Express \cite{Shafi2009}, FastMPJ \cite{Taboada2012},
and Open MPI Java \cite{openmpiJava}. In this context, we choose
MPJ Express that is being developed and maintained at NUST Pakistan.
For the third and the final section of the course we choose the Apache
Hadoop software \cite{ApacheHadoop} for implementing MapReduce applications, which is
a popular open-source software to process---through automatic
parallelization---large amounts of data.

Rest of this paper is organized as follows. Section \ref{sec:Examples} introduces
HPC workloads used as sample applications throughout the course. Section \ref{sec:courseContent}
outlines the course syllabus followed by a detailed discussion on course contents. Section \ref{sec:conclusion} concludes the paper.

%\section{Motivation of Java}
%\label{sec:MotivationOfJava}

\section{HPC Workloads}
\label{sec:Examples}

This section introduces HPC workloads used as sample applications throughout the course.
We first introduce sequential version of these applications. Students were invited to develop
shared and distributed memory versions using threads and messaging in the first and second
sections of the course.

Typically parallel computations can be roughly divided into two categories
based on their requirements to communicate during the computation phase.
Applications that do not require any communication in computation phase are
called \emph{embarrassingly parallel} computations, while others that require
periodic communication in computation phase are generally referred to as
\emph{synchronous} computations. We choose three embarrassingly parallel
computations, which included $\pi$ Calculation, Mandelbrot Set Calculation, and Matrix-Matrix
Multiplication. Also, we choose two synchronous computations that included
Conway's Game of Life and Laplace Equation Solver. Rest of the section presents
an overview of each of the sample application.

\subsection{$\pi$ Calculation}
%\emph{\textbf{$\pi$ Calculation:}}
%\label{sec:piCalculation}

There are many ways to approximately calculate the mathematical constant $\pi$. One ``brute force'' method is based on the following formula:
\begin{displaymath}
\pi \approx h\sum_{i=0}^{N-1} \frac{4}{1 + (h(i+\frac{1}{2}))^2}
\end{displaymath}
where $N$ is the number of steps and $h$ is the size of a single step. We set $N$ to $10$ million, which is sufficiently large to get an accurate estimate of $\pi$. The above formula for calculating $\pi$ is simply a large sum of independent terms. Listing \ref{list:piCal} shows the sequential version of the $\pi$ calculation code.

\begin{lstlisting}[caption=Serial $\pi$ Calculation Code, label={list:piCal}]
for(int i=0; i<numSteps; i++) {
  double x=(i+0.5) * step;
  sum += 4.0/(1.0 + x*x);
}
double pi=step * sum ;
\end{lstlisting}
Calculation of mathematical constant $\pi$ is an embarrassingly parallel application and it provides good starting point in learning parallel programming techniques.

\subsection{The Mandelbrot Set}
%\emph{\textbf{Mandelbrot Set:}}
%\label{sec:mandelbrotset}

\begin{figure}
\centering     %%% not \center
\subfigure[Mandelbrot Set]{\label{fig:mandelbrotSet}
\includegraphics[width=0.21\textwidth]{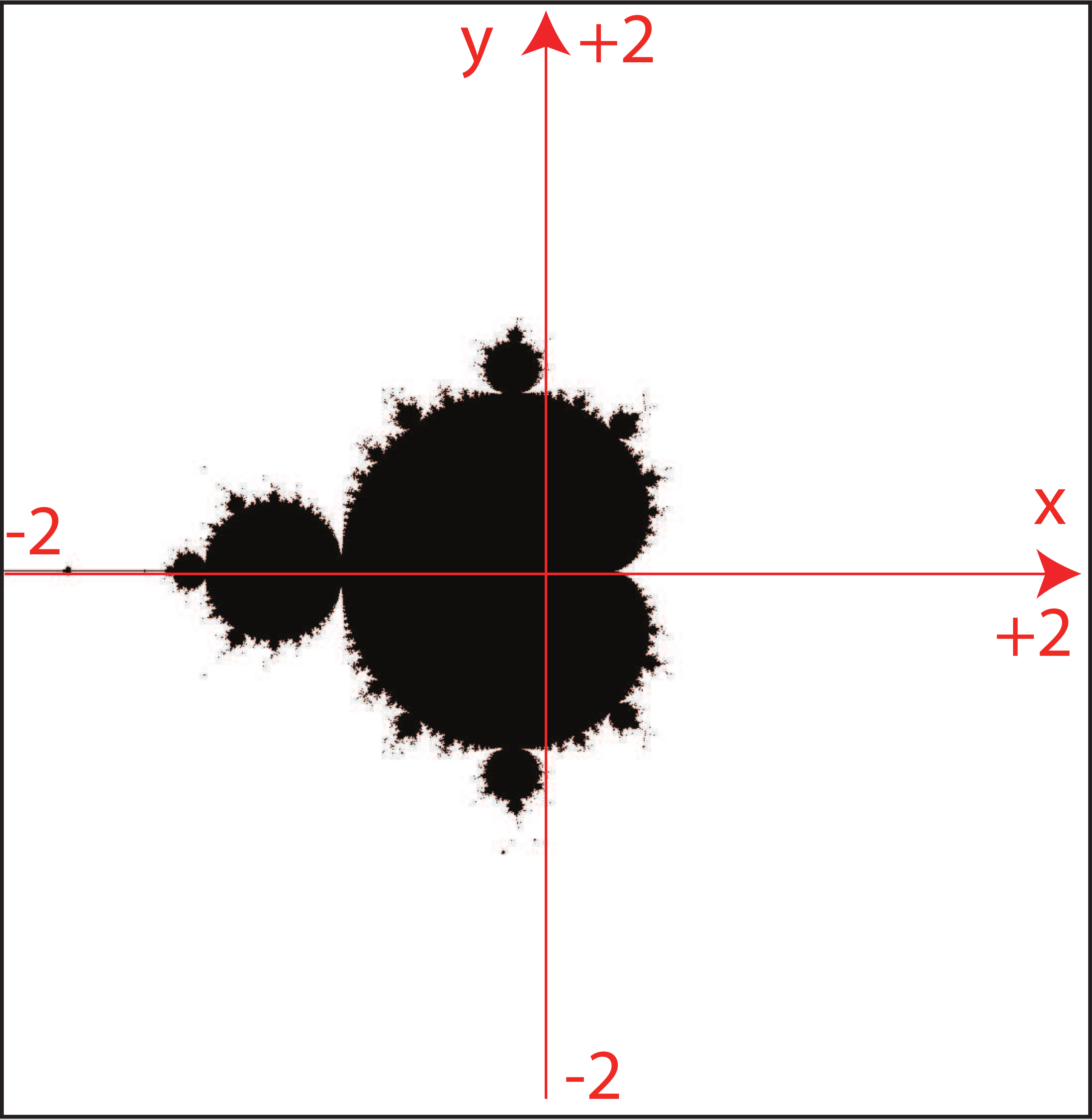}}
\subfigure[Convay's Game of life]{\label{fig:Gameoflife}
\includegraphics[width=0.21\textwidth]{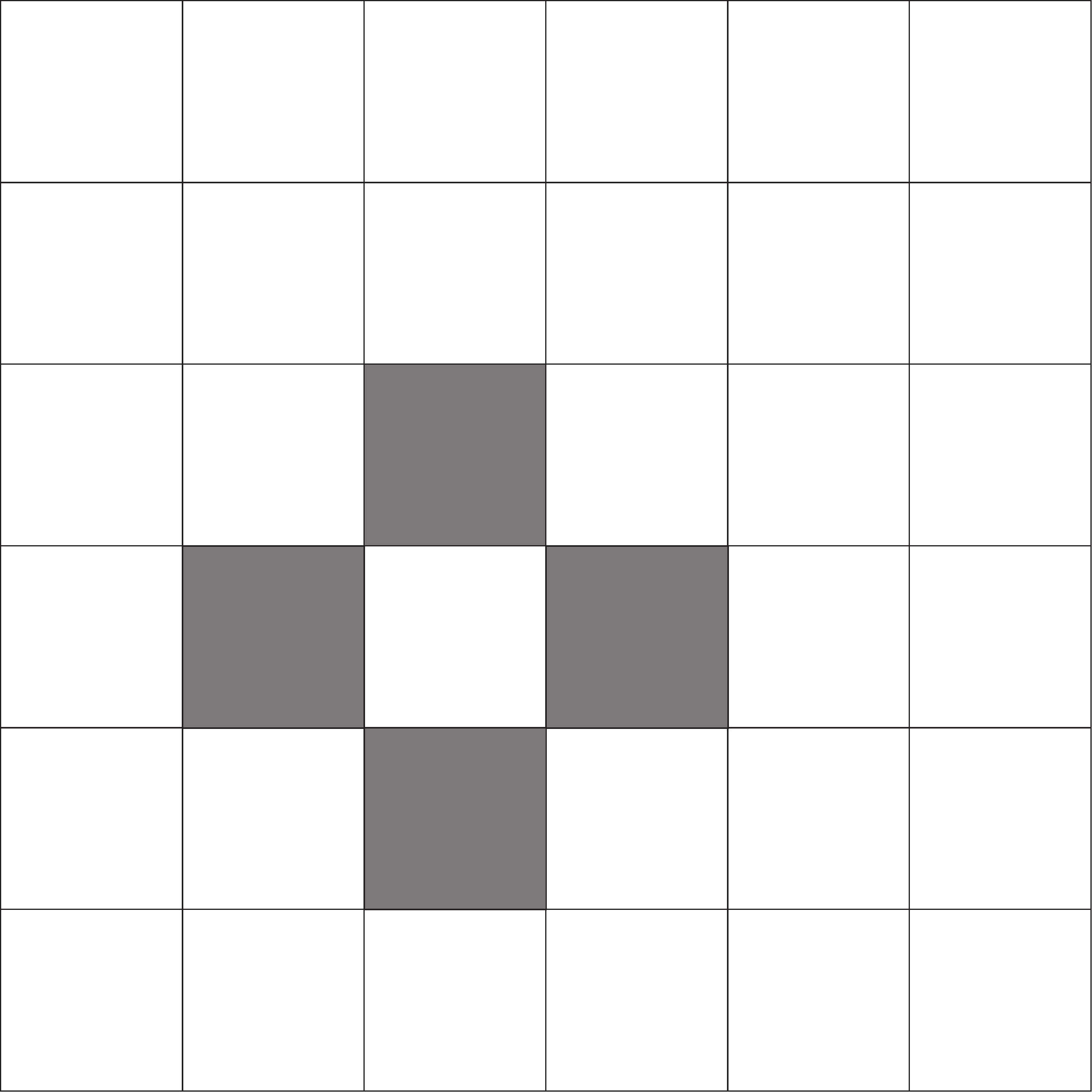}}
\caption{HPC Workloads}
\end{figure}

The Mandelbrot Set is a collection of complex numbers that are
quasi-stable---values increase and decrease but do not exceed
a particular limit. The set is computed by iterating the function:
\begin{displaymath}
z_{k+1} = z_{k}^{2} + c
\end{displaymath}
Iterations continue until magnitude of $z$ is greater than $2$, or
the number of iterations reaches arbitrary limit. The Mandelbrot Set
can be seen in Figure \ref{fig:mandelbrotSet} where mathematically
the set is the black area within part of $x,y$ plane
with $ -2 \leq x,y \leq 2 $.

Listing \ref{list:mandelbrotSet} depicts the serial pseudo code
for the Mandelbrot Set. As shown, the innermost while loop repeats forever
if we are in the black region; in practice, stop the loop after some {\tt CUTOFF} number of iterations.

\begin{lstlisting}[caption=Serial Mandelbrot Set Calculation Code, label={list:mandelbrotSet}]
for(int i=0; i<N; i++) {
  for(int j=0; j<N; j++) {
    double x=step * i - 2.0; //  -2<=x<=2
    double y=step * j - 2.0; //  -2<=y<=2
    complex c=(x, y), z=c;
    int k = 0;
    while (k<CUTOFF && abs(z)<2.0) {
      z=c + z * z;
      k++ ;
    }
    set[i][j]=k;
  }
}
\end{lstlisting}

\subsection{Matrix Multiplication}
%\emph{\textbf{Matrix Multiplication:}}
%\label{sec:matrixmultiplication}

Another embarrassingly parallel computation discussed in the course
included the Matrix-Matrix and Matrix-Vector multiplication. It is a useful
scientific kernel where parallelism not only helps in improving code performance
but also allows solving larger matrices on parallel hardware. Elements of the
resultant matrix $C$ are produced by multiplying matrices $A$ and $B$ as follows:

\begin{displaymath}
C_{i,j} = \sum_{k=0}^{l-1} A_{i,k} B_{k,j}
\end{displaymath}

Listing \ref{list:matrixMulti} shows pseudo code for multiplying two square
matrices $A$ and $B$.

\begin{lstlisting}[caption=Serial Matrix Multiplication Code, label={list:matrixMulti}]
for (i=0; i<n; i++) {// for A's rows
  for (j=0; j<n; j++) {// for B's columns
    c[i][j] = 0;
    for (k = 0; k < n; k++)
      c[i][j]=c[i][j]+a[i][k]*b[k][j];
  }
}
\end{lstlisting}

\subsection{Conway's Game of Life}
%\emph{\textbf{Conway's Game of Life:}}
%\label{sec:gameoflife}

Conway's Game of Life is a cellular automaton on a $2D$ grid as shown in Figure \ref{fig:Gameoflife}, where each cell takes the value $0$ (dead) or $1$ (alive). As part of the simulation, newer generations of cells are evolved according to a pre-defined criteria. At each timestep, the new value of each cell depends on its old value and old values of the neighbouring cells. Listing \ref{list:gameoflife} shows the pseudo code for the sequential version
of the Game of Life. The {\tt cells} array is the main data-structure while the auxiliary array {\tt sums} holds the sum of cell elements neighbouring the cell $(i,j)$ after the sum phase. As the {\tt while} loop executes, cell values---stored in the {\tt cells} array---evolve from one generation to another.

\begin{lstlisting}[caption=Serial Conway's Game of Life Code, label={list:gameoflife}]
while(true) {
  // Sum Phase
  for(int i=0; i<N; i++)
    for(int j=0; j<N; j++)
      sums[i][j]=sum of cells values neighbouring (i, j);
  // Update Phase
  for(int i=0; i<N; i++)
    for(int j=0; j<N; j++)
      cells[i][j]=update(cells[i][j], sums[i][j]);
}
\end{lstlisting}

\subsection{Laplace Equation Solver}
%\emph{\textbf{Laplace Equation:}}
%\label{sec:laplaceequation}

The two-dimensional Laplace equation is an equation that crops up in
several places in physics and mathematics. We choose Laplace equation
as a sample application in this course because it is a relatively
simple numerical problem---in science and engineering---that can be
tackled by parallel programming. The {\em discrete} version of the
Laplace equation on a two-dimensional grid of points can be stated as:

\begin{displaymath}
\nabla^{2}u = \frac{\partial^{2}u}{\partial{x^{2}}} + \frac{\partial^{2}u}{\partial{y^{2}}} = 0
\end{displaymath}

Listing \ref{list:laplaceEq} shows the sequential code for solving Laplace equation.
Here the main data-structure is the {\tt phi} array, which stores unknown variables
of the equation as its elements. An {\em iterative} numerical
approach to solving the equation is just to initially set all elements of {\tt phi}
that we have to solve for to some value like zero, then repeatedly change
individual {\tt phi [i][j]} elements to be the average of their neighbours.
If we repeat this local update sufficiently many times, the {\tt phi} elements
converge to the global solution of the equations. This is called the
{\em relaxation method}.

\begin{lstlisting}[caption=Serial Laplace Equation Solver, label={list:laplaceEq}]
for(int iter=0; iter<NITER; iter++) {
  // Calculate new phi
  for(int i=1; i<(N-1); i++) {
    for(int j=1; j<(N-1); j++) {
      phi[i][j] = 0.25F * (phi[i][j-1] + phi[i][j+1] + phi[i-1][j] + phi[i+1][j]);
    }
  }
}
\end{lstlisting}

\begin{figure}[h]
  \center
  \includegraphics [width=0.21\textwidth]{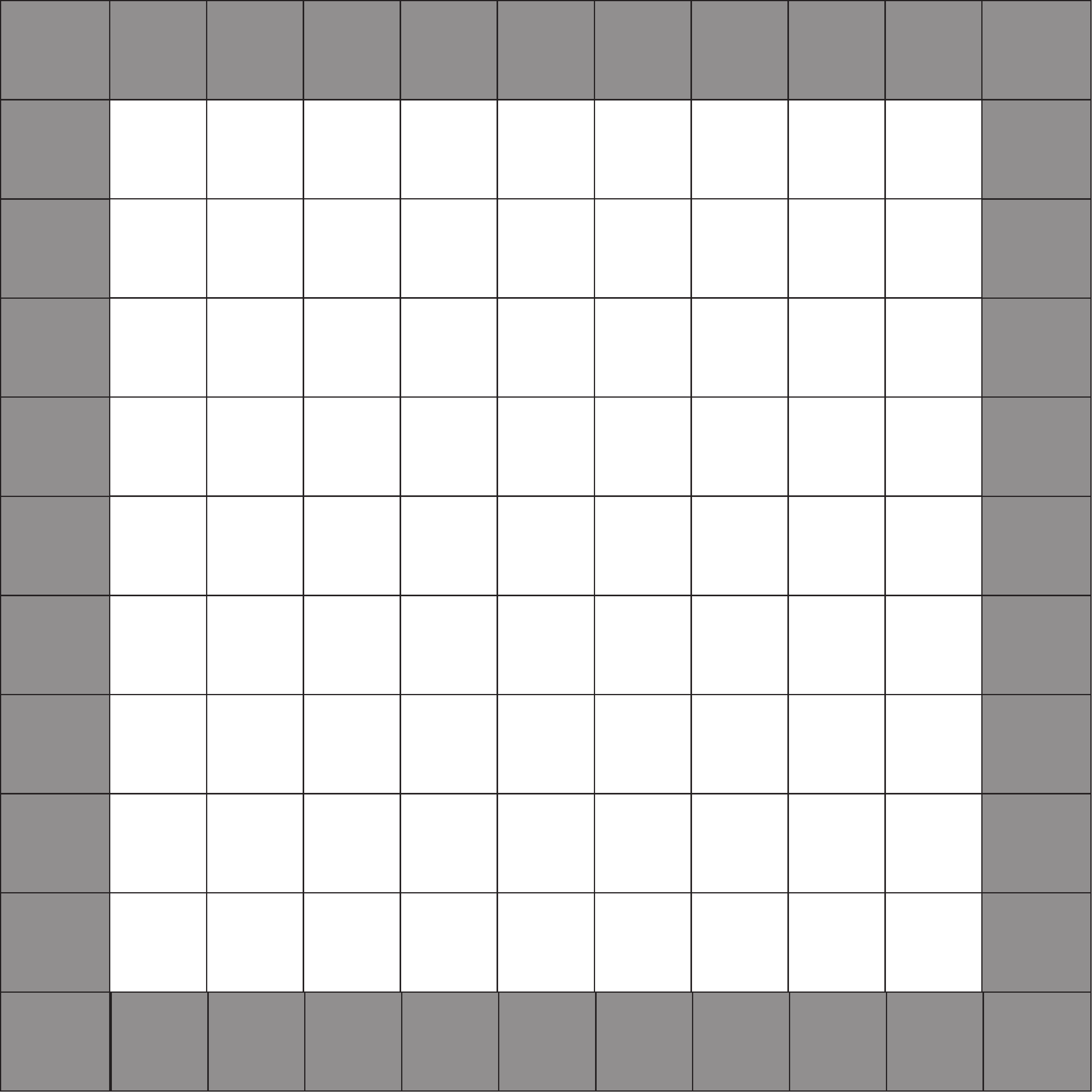}
  \caption{Laplace Equation Solver Grid Points}
  \label{fig:laplaceEq}
\end{figure}

\section{Course Contents}
\label{sec:courseContent}

%This section is divided into three sub-sections. First sub-section outline programming techniques for shared memory systems. Second sub-section describe programming tools and APIs for distributed memory systems followed by discussion on advance topics in third sub-section.
This section discusses details of course contents taught in this course.
We begin by presenting an overview of the course syllabus. This is followed
by covering shared and distributed memory parallel programming techniques.
The section concludes with a discussion on advanced topics, which were part
of this course.

%Introductory lectures started with motivation of high performance computing and its usage in various applications, top $3$ supercomputers from the list of top500 supercomputers\footnote{http://www.top500.org/list/2013/11/} followed by HPC facilities available at SEECS-NUST. Some important concepts including  shared/distributed  memory systems, performance measurement
% metrics, hardware accelerators, emergence of multicore processors and Moore's law concluded initial lectures.  After initial introduction, course was divided into three sections namely \emph{shared memory programming}, \emph{distributed memory programming} and \emph{advance topics}. Details of each section is given below.

\subsection{Course Syllabus}
\label{sec:syllabus}

This sub-section presents an overview of the course syllabus.
The weekly distribution of lectures, labs, and assignments is shown in
Table \ref{tab:syllabusTable}. The duration of the course was eighteen
weeks. There are two One Hour Tests (OHTs) in week six and twelve.
In addition, there is one End Semester Exam (ESE) that takes place
in the last week, that is the eighteenth week. This was a $2+1$ credit
hours course, which implies two one hour weekly lectures and one lab.
There are three contact hours for the weekly lab. The shared memory
part of the course was covered in the first four weeks followed by the
first OHT in week six. Similarly, the distributed memory part of the
course was covered from week five to eleven followed by the second
OHT in week twelve. The last week teaching weeks---from thirteen to
seventeen---covered advanced topics followed by the ESE in week
eighteen.

%This sub-section presents an overview of syllabus of parallel computing course. Week wise distribution of syllabus is outlined followed by weightage of theoretical and practical parts.

%We divided our course contents into three broad sections namely \emph{shared memory programming}, \emph{distributed memory programming} and \emph{advance topics}. The shared memory programming section covered programming techniques for shared memory
%systems including Java threads \cite{threadsJava}, OpenMP \cite{openmp}, and Intel
%Cilk Plus \cite{cilkplus}. category consisted of two parts. The second section covered programming tools  and APIs for distributed memory systems including commodity clusters. For this purpose we used Java MPI library, MPJ Express. The advance topics section covered contemporary topics in HPC including lectures on GPU programming using CUDA and MapReduce programming models using Hadoop.
%Week wise distribution of course syllabus is given in Table \ref{tab:syllabusTable}. %where lectures conducted in first five weeks covered shared memory programming followed by OHT1 in 6th week; lectures conducted during five weeks after OHT1 covered distributed memory programming followed by OHT2 in 12th week. Lectures covered from OHT2 to final exam comprised of advance topics.

\begin{table}[!h]
\centering
\caption{Parallel Programming Course Syllabus}
\label{tab:syllabusTable}
\begin{tabular}{|m{0.06\textwidth}|m{0.37\textwidth}|} \hline
%\begin{tabular}{|l|p{'5cm'}|} \hline
\multicolumn{1}{|c}{\cellcolor[gray]{0.8}{\bf Week}} & \multicolumn{1}{|c|}{\cellcolor[gray]{0.8}{\bf Topic}}\\ \hline
\multirow{3}{*}{Week-1} & Introduction to Parallel Computing \\
                        & Review of Java Threads \\
                        & Lab 1: $\pi$ Calculation \\ \hline
\multirow{3}{*}{Week-2} & Introduction to Parallel Hardware \\
                        & Parallel Programming Approaches \\
                        & Lab 2: Array Operations \\ \hline
\multirow{4}{*}{Week-3} & Embarrassingly Parallel Computations \\
                        & Shared Memory Programming \\
                        & Lab 3: Mandelbrot Set Calculation \\
                        & Assignment 1: Monte Carlo $\pi$ calculation \\ \hline
\multirow{3}{*}{Week-4} & Introduction to OpenMP \\
                        & Introduction to Cilk \\
                        & Lab 4: Parallelizing Game Of Life  \\ \hline
\multirow{4}{*}{Week-5} & Distributed Memory Systems and MPI \\
                        & Data Decomposition and MPI Communication \\
                        & Lab 5: Solving Laplace Equation \\
                        & Assignment 2: Dense matrix multiplication \\  \hline
\multicolumn{2}{|c|}{\cellcolor[gray]{0.7}One Hour Test-1}\\ \hline
\multirow{2}{*}{Week-7} & Features of MPI \\
                        & Lab 6: $\pi$ Calculation using MPI \\ \hline
\multirow{2}{*}{Week-8} & MPJ Express Programming \\
                        & Lab 7: Mandelbrot Set calculation using MPI \\ \hline
\multirow{3}{*}{Week-9} & Global and Local Synchronization \\
                        & Lab 8: Solving Game of life using MPI \\
                        & Assignment 3: Monte Carlo $\pi$ calculation using MPI \\ \hline
\multirow{2}{*}{Week-10} & MPI Point to Point Communication \\
                        & Lab 9: Array Operations using MPI  \\ \hline
\multirow{2}{*}{Week-11} & MPI Collective Communication \\
                        &  Lab 10: Solving Laplace Equation using MPI  \\ \hline
\multicolumn{2}{|c|}{\cellcolor[gray]{0.7}One Hour Test-2}\\ \hline
\multirow{2}{*}{Week-13} & GPU Programming - I \\
                        &  Lab 11: Mandelbrot Set using MPI Collectives  \\ \hline
\multirow{2}{*}{Week-14} & GPU Programming -II \\
                        & Lab 12: N-body Simulations \\ \hline
\multirow{2}{*}{Week-15} & Motivation of MapReduce\\
                        & Lab 13: Array operations using GPUs \\ \hline
\multirow{2}{*}{Week-16} & Apache Hadoop\\
                        & Lab 14: Word count using MapReduce \\ \hline
\multirow{1}{*}{Week-17} & Course Review  \\ \hline
\multicolumn{2}{|c|}{\cellcolor[gray]{0.7}End Semester Exam}\\ \hline
\end{tabular}
\end{table}

Table \ref{tab:gradingWeightage} shows our grading policy clearly depicting
the weightage assigned to theoretical and practical parts of the course.
Weekly labs were conducting in a typical teaching lab, which included forty
PCs connected via Gigabit Ethernet to one another. Each PC comprised of
Intel\textsuperscript{\textregistered} Core\textsuperscript{\texttrademark} i5-3470 CPU
and 4 GBytes of main memory.

\begin{table}[!h]
\centering
\caption{Grading}
\label{tab:gradingWeightage}
\begin{tabular}{|m{0.16\textwidth}|c|m{0.16\textwidth}|c|} \hline
%\begin{tabular}{|l|p{'5cm'}|} \hline
\multicolumn{2}{||c|}{\cellcolor[gray]{0.8}{\bf Theoretical (70\%)}} & \multicolumn{2}{|c|}{\cellcolor[gray]{0.8}{\bf Practical (30\%)}}\\ \hline

One Hour Tests & 35\% & \multirow{2}{*}{Weekly labs} & \multirow{2}{*}{80\%}\\ \cline{1-2}
End Semester Exam & 45\% & & \\ \hline
Quizzes & 15\% & \multirow{2}{*}{Lab Exam} & \multirow{2}{*}{20\%}\\ \cline{1-2}
Assignments & 5\% & & \\ \hline

\end{tabular}
\end{table}

\subsection{Shared Memory Parallel Programming}

This sub-section outlines shared memory parallel programming techniques
covered in this course. In this context, the course reviewed the Java
threads API as a viable option for writing shared memory concurrent
programs. Lectures covered as part of this section demonstrated using
threads for implementing embarrassingly parallel and synchronous
computations introduced earlier in Section \ref{sec:Examples}. While
covering these HPC workloads, the instructor illustrated important
parallelism concepts including problem decomposition/partitioning,
load balancing, and synchronization. Two main partitioning
techniques namely block-wise and cyclic distributions were covered---
see Listing \ref{list:dd} for code patterns for the two distributions.

\begin{lstlisting}[caption=Decomposition using Block/Cyclic Distributions, label={list:dd}]
// Original for loop
for (int i=0; i<N; i++)}
// me=current thread; P=total threads
// block-wise distribution of each thread
for (int i=me*N/P; i<(me+1)*N/P; i++)
// cyclic distribution of each thread
for (int i=me; i<N; i+=P)
\end{lstlisting}

\begin{figure}
\centering     %%% not \center
\subfigure[Vertically]{\label{fig:mandelbrotSetVertically}
\includegraphics[width=0.19\textwidth]{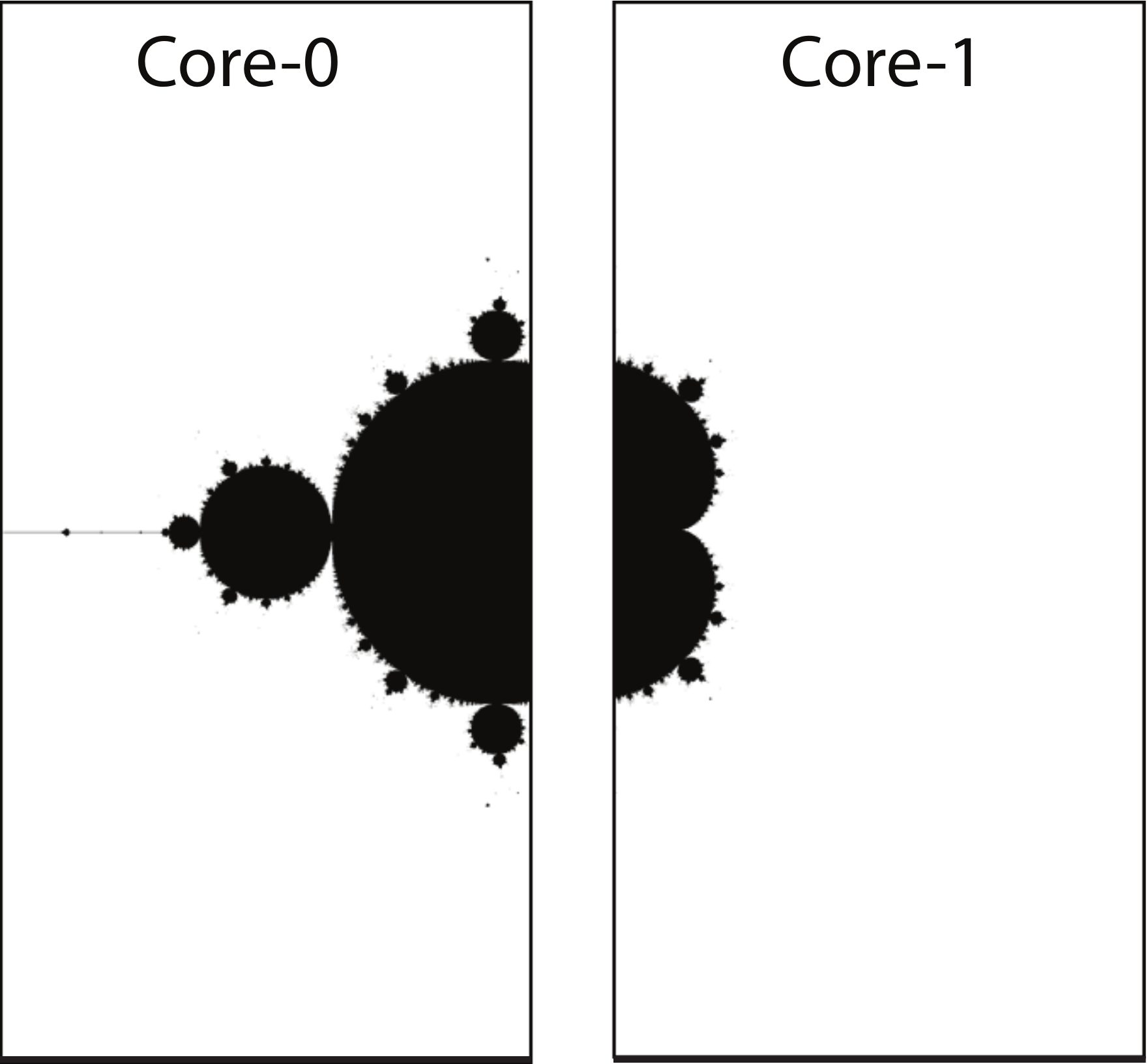}}
\subfigure[Horizontally]{\label{fig:mandelbrotSetHorizontally}
\includegraphics[width=0.22\textwidth]{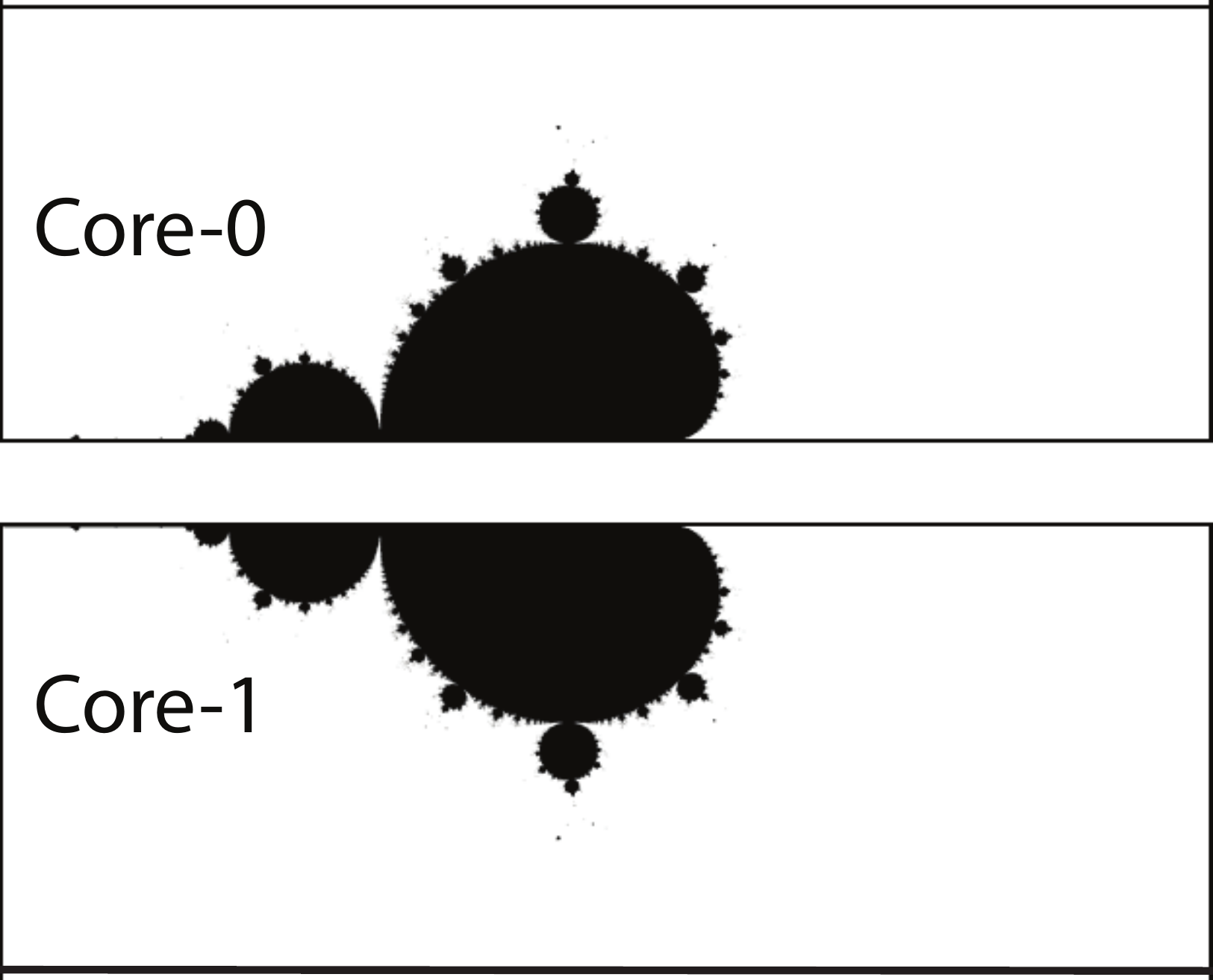}}
\caption{Partitioning the $(x,y)$ plane}
\end{figure}

\begin{figure}[t]
  \center
  \includegraphics [width=0.21\textwidth]{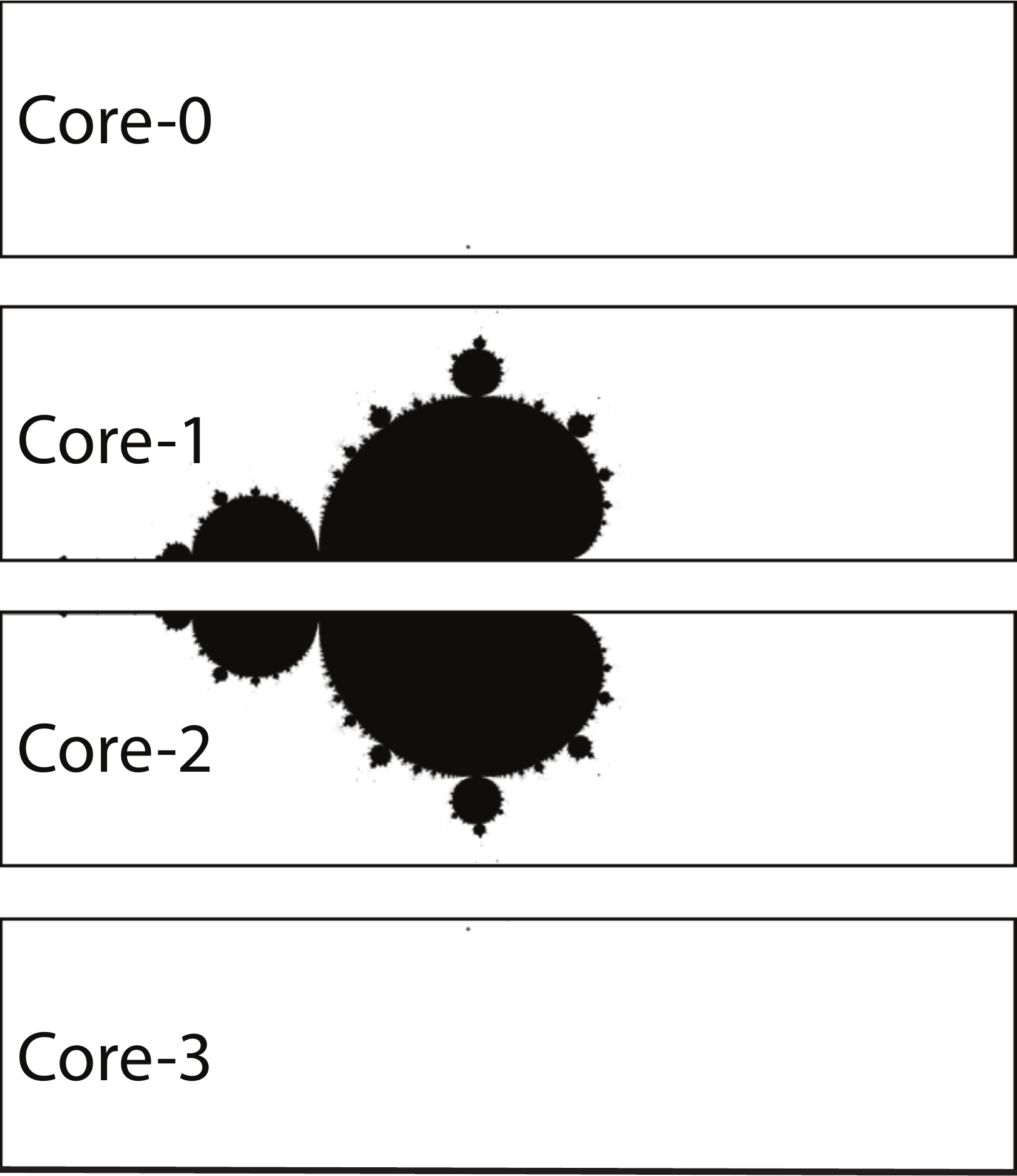}
  \caption{Partitioning the $(x,y)$ plane in four horizontal blocks}
  \label{fig:mandelbrotSet4halves}
\end{figure}

\textbf{The Mandelbrot Set:}
As part of our coverage on parallelizing embarrassingly parallel computations
using Java threads on shared memory platforms, there was a discussion on
multicore-enabling the Mandelbrot Set code. This dialog also demonstrated
key topics including partitioning and load balancing. Our initial attempts to
develop a multi-threaded Mandelbrot Set calculation code were based on dividing
the $(x,y)$ plane into two halves both vertically and horizontally. Poor load
balancing was observed in the vertical division as shown in
Figure \ref{fig:mandelbrotSetVertically}. The reason is that core 0
had more substantially work than core 1. By dividing the $(x,y)$ plane
horizontally into two halves, we noted perfect load balancing due to
symmetry of the Mandelbrot Set---this can be seen in
Figure \ref{fig:mandelbrotSetHorizontally}. However, if the horizontal
partitioning is carried out on four cores, then we also observed
poor load balancing as core 0 and 3 have little fraction of the
total computational work---this is depicted in
Figure \ref{fig:mandelbrotSet4halves}. It was discussed during lectures
that this particular issue can be tackled by using cyclic distribution,
which is difficult to implement and sometimes less efficient due to poor
usage of cache but also has merits in certain applications like the
Mandelbrot Set calculation.

%\begin{figure}
%\centering     %%% not \center
%\subfigure[Dividing x, y plane horizontally in 4 halves]{\label{fig:mandelbrotSet4halves}
%\includegraphics[width=0.21\textwidth]{Images/mandelbrotSet4halves.png}}
%\subfigure[Partitioning in Image processing]{\label{fig:imageProcessing}
%\epsfig{file=imageProcessing.eps, width=0.21\textwidth}}
%\includegraphics[width=0.21\textwidth]{Images/imageProcessing.png}}
%\caption{Partitioning in embarrassingly parallel problems}
%\end{figure}

%\textbf{Image Processing:}
%Low level image processing operations such as shifting, scaling and rotating are notable embarrassingly parallel problems.  Students were demonstrated how to perform load balancing in image processing by partitioning image into regions for individual processes (see Figure \ref{fig:imageProcessing}.

\textbf{Matrix Multiplication:}
Parallelizing Matrix-Matrix and Matrix-Vector computations with Java threads
was discussed next. For dense matrices, it is possible to achieve
good speedups by exploiting traditional vertical block-wise partitioning.
However, we noted that this traditional partitioning strategy does not
work well for sparse matrices. In addition, sparse matrices unnecessarily
waste memory if stored in two-dimensional array format due to large
numbers of zeroes. This is solved by utilizing a special data-structure
that keeps track of row, column, and value of each non-zero element in the
matrix. This information can be stored as {\em triples} in the form of an
array list or a linked list. Parallelization can now be achieved by partitioning
the list of triples instead of the original sparse matrix.

\begin{figure}[t]
  %\center
  %\epsfig{file=gameoflifeRaceCondition.eps, width=0.47\textwidth}
  \includegraphics [width=0.47\textwidth]{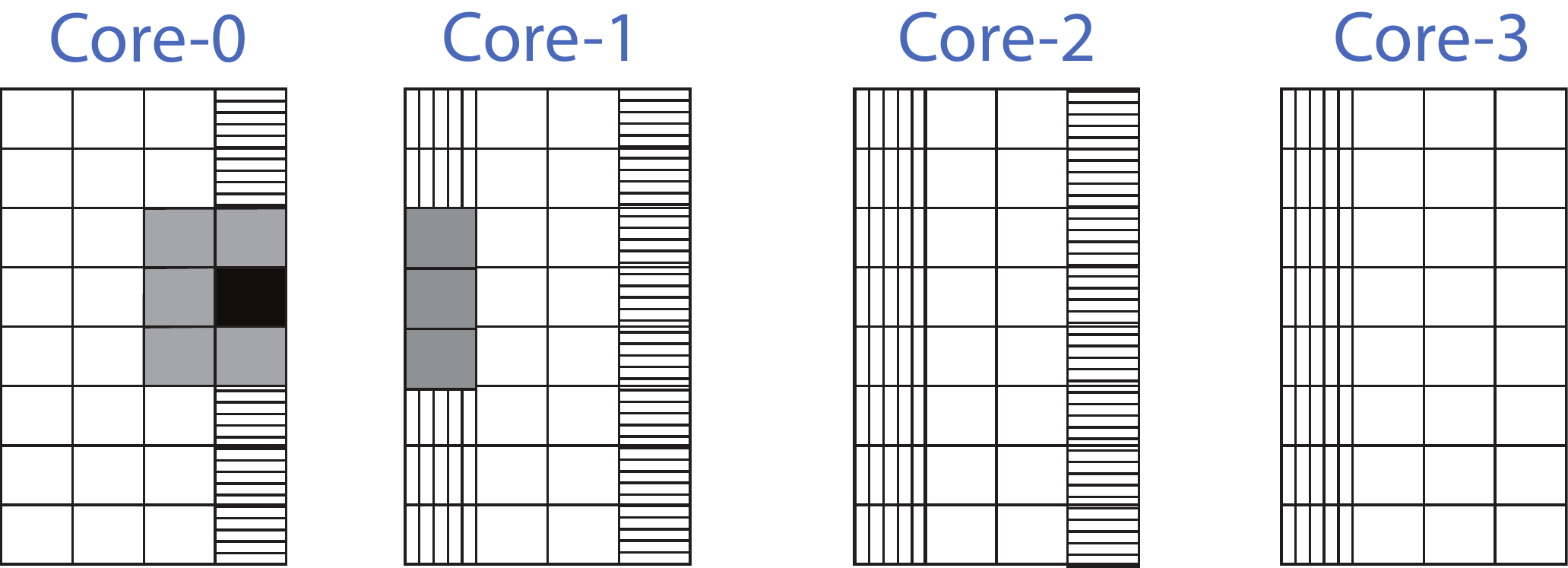}
  \caption{Partitioning in Multi-threaded Conway's Game of Life}
  \label{fig:gameoflifeRaceCondition}
\end{figure}

\textbf{Conway's Game of life:}
 As part of our coverage on parallelizing synchronous computations using
 Java threads, we started off with Conway's Game of Life. The discussion began
 with a review of the sequential code---shown in Listing \ref{list:gameoflife}---that
 includes two computational phases called sum and update. In the sum phase,
 the code calculates sums of all neighbours of {\tt cells[i][j]}---this sum is
 stored in {\tt sums[i][j]}. In the update phase, the new value is written to
 {\tt cells[i][j]}. In our first attempt, we developed a threaded version
 that used conventional block-wise partitioning strategy to execute sum and update
 phases concurrently in multiple threads. However, this version introduced
 a race condition in the code that resulted in an unpredictable behavior. The
 reason was that some threads ran ahead of other threads perhaps by several
 generations due to lack on any synchronization/communication between
 concurrent threads. This issue is depicted in Figure \ref{fig:gameoflifeRaceCondition}
 where core 0 is writing the border black cell (in the update phase) and while
 doing so, it is also reading all grey cells (in the sum phase). In general, all cells with
 vertical stripes are written by the {\em owner} thread and read by the
 left neighbour. Similarly all cells with horizontal stripes are
 written by the owner thread and read by the right neighbour thread. This issue was
 tackled by employing {\em barrier synchronization}. For this purpose, the parallel
 code instantiated an object of the {\tt java.util.CyclicBarrier} class. This object
 was used to call the {\tt await()} in co-operating threads to achieve barrier
 synchronization. This function call blocks until all $P$ threads---taking part
 in the computation---have made this call. Parallelized version of Conway's Game of
 Life with barrier synchronization can be seen in Listing \ref{list:gameoflifeMulti}.

\begin{lstlisting}[caption=Multi-threaded Conway's Game of Life Code, label={list:gameoflifeMulti}]
Class LifeThread {
  void run() {
    while(true) {
      // Sum Phase
      for(int i=begin; i<end; i++)
        for(int j=0; j<N; j++)
          sums[i][j]=sum of cells values neighbouring(i, j);
      barrier.await() ;
      // Update Phase
      for(int i=begin; i<end; i++)
        for(int j=0; j<N; j++)
          cells[i][j]=update(cells[i][j], sums[i][j]);
      barrier.await() ;
    }
  }
}
\end{lstlisting}

\subsection{Distributed Memory Parallel Programming}

This sub-section outlines distributed memory parallel programming techniques
covered in this course. In this context we began by reviewing MPJ Express---a
Java MPI library. Most parallel programs designed to run on large clusters
utilize MPI for messaging. We noted that MPI implements the Single Program
Multiple Data (SPMD) model---each process runs the same program, but of course
operates on its own local memory (data). Lectures covered as part of this
section demonstrated using messaging---as provided by MPJ Express---for
implementing embarrassingly parallel and synchronous
computations introduced earlier in Section \ref{sec:Examples}. While
covering these HPC workloads, we reviewed and practiced using point-to-point and
collective communication routines provided by the MPJ Express software.

%Previous lectures concentrated on parallel programming with shared memory. But due to limited scalability of shared memory systems we are forced to consider distributed memory systems.
%A group of ordinary workstations connected by a network (Ethernet, Myrinet, Infiniband) is perhaps the most basic instantiation of distributed memory systems.
%\begin{figure}[t]
%  \center
%  \includegraphics [width=0.41\textwidth]{Images/distributedCluster.png}
%  \caption{A Simple Cluster}
%  \label{fig:distributedCluster}
%\end{figure}
%A new concept \emph{cooperating processes} was introduced to students. Biggest change in programming distributed memory systems is that we move from \emph{cooperating threads}---sharing access to a common memory---to \emph{cooperating processes}, which each have their own private memory, but no with memory shared between processes. In some sense the lowest common denominator approach, explicit Message Passing is widely used in distributed memory parallel programs today.
 %one particular API (Application Programming Interface) for this. Twenty years ago, \emph{Message Passing Interface} (MPI) version $1$ was standardised by a group of manufacturers and academics called the \emph{MPI Forum}.

%We used an unofficial Java API for teaching MPI, as implemented in the software called ``MPJ Express''.
Listing \ref{list:mpjHelloWorld} shows the most basic MPJ Express program.
The MPJ Express library is initialized and finalized using the {\tt MPI.Init(args)}
and {\tt MPI.Finalize()}, respectively. Once initialized, the MPI library provides
access to a special data-structure called {\em communicator}, which encapsulates
all processes taking part in the parallel execution. In our code this data-structure
is represented by the {\tt MPI.COMM\_WORLD} object, which can be used for calling various
routines including querying total number of parallel processes---via
{\tt MPI.COMM\_WORLD.Size()}---and a process' own {\em rank}---via {\tt MPI.COMM\_WORLD.Rank()}.
The {\tt MPI.COMM\_WORLD} object can also be used for invoking communication routines
including point-to-point and collective communication. Listing \ref{list:sendrecv} shows
signatures for the most basic blocking send and receive primitives provided by MPJ Express.

\begin{lstlisting}[caption=MPJ Express Hello World Code, label={list:mpjHelloWorld}]
import mpi.*;
public class HelloWorld {
  public static void main(String args[]) throws Exception {
    MPI.Init(args) ;
    // Get total number of processes
    int P = MPI.COMM_WORLD.Size();
    // Get rank of each process
    int me = MPI.COMM_WORLD.Rank();
    MPI.Finalize() ;
  }
}
\end{lstlisting}

\begin{lstlisting}[caption=MPJ Express Send and Receive Methods, label={list:sendrecv}]
void Comm.Send(Object buf, int offset, int count, Datatype type, int dest, int tag)
Status Comm.Recv(Object buf, int offset, int count, Datatype type, int src, int tag)
\end{lstlisting}

\textbf{$\pi$ Calculation:} We started our discussion of HPC workloads with the classic
embarrassingly parallel $\pi$ calculation code. Listing \ref{list:piCalcMPI} shows
the sketch of the MPI version of the $\pi$ calculation code. This code utilizes the
primitive messaging functions like {\tt Send()} and {\tt Recv()}. Each MPI process
independently calculates its own contribution to the {\tt sum} variable, which is
communicated to the master process (rank 0). Once the master process has received
all contributions from {\em slave/worker} processes, it calculates the final sum
that is used to generate the value of $\pi$.

\begin{lstlisting}[caption=$\pi$ Calculation using MPJ Express, label={list:piCalcMPI}]
if (rank != 0) {
  double[] sendBuf=new double[]{sum};
  //1-element array containing sum
  MPI.COMM_WORLD.Send(sendBuf, 0, 1, MPI.DOUBLE, 0, 10);
}
else { //rank == 0
  double[] recvBuf=new double[1] ;
  for (int src=1 ; src<P; src++) {
    MPI.COMM_WORLD.Recv(recvBuf, 0, 1, MPI.DOUBLE, src, 10);
    sum += recvBuf [0] ;
  }
}
double pi = step * sum ;
\end{lstlisting}

\textbf{The Mandelbrot Set:}
Another embarrassingly parallel computation discussed in the class
was the Mandelbrot Set. Here, students were invited to develop a
distributed memory version of the Mandelbrot Set code using similar
partitioning and communication patterns used in the $\pi$ calculation code.
Also, non-blocking (asynchronous) communication primitives that allow communication/computation
overlap were introduced. Table \ref{tab:p2pCommunication} summarizes
the blocking and non-blocking point-to-point communication routines
provided by MPJ Express. Another utility function {\tt Sendrecv()} was
introduced, which essentially combines the {\tt Send()} and {\tt Recv()}
functionality in a single call.

\begin{table}[!h]
\centering
\caption{Point-To-Point Communication Modes}
\label{tab:p2pCommunication}
\begin{tabular}{|m{0.21\textwidth}|m{0.21\textwidth}|} \hline
%\begin{tabular}{|l|p{'5cm'}|} \hline
\multicolumn{1}{|c|}{\cellcolor[gray]{0.8}{\bf Blocking}} & \multicolumn{1}{|c|}{\cellcolor[gray]{0.8}{\bf Non-blocking}}\\ \hline
{\tt Send} & {\tt Isend} \\ \hline
{\tt Recv} & {\tt Irecv} \\ \hline
{\tt Bsend} (Buffered) & {\tt Ibsend} (Buffered)\\ \hline
{\tt Ssend} (Synchronous) & {\tt Issend} (Synchronous)\\ \hline
{\tt Rsend} (Ready)& {\tt Irsend} (Ready)\\ \hline
\end{tabular}
\end{table}

\begin{figure}[t]
  \center
  \includegraphics [width=0.47\textwidth]{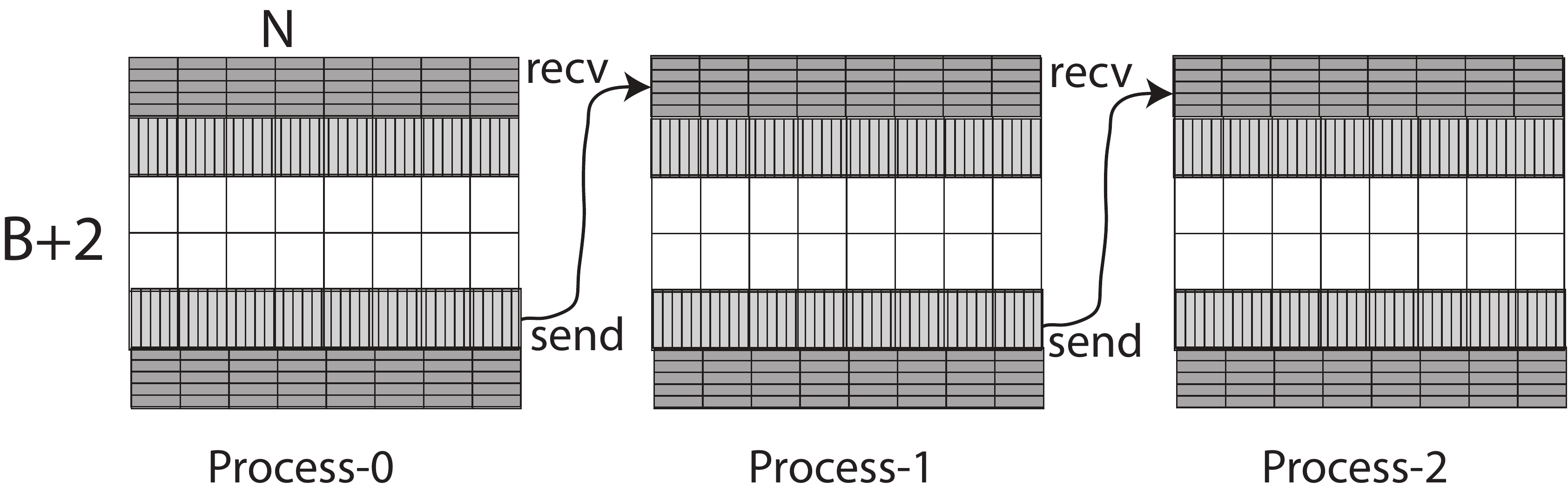}
  \caption{Partitioning in MPI Version of Conway's Game of Life}
  \label{fig:ghostRegions}
\end{figure}

 \textbf{Conway's Game of life:} A distributed memory version of the Conway's
 Game of Life code was also discussed in the class. Once partitioning has been
 performed, each of the MPI process executes the sum and update phases in a concurrent
 fashion. However communication needs to take place during the sum phase for all the
 border cells belonging to the top and bottom row. Performing explicit communication
 for each cell during the sum phase is an expensive operation. This can be optimized
 by exchanging border rows amongst neighbouring processes before an MPI process enters the
 sum phase. To implement this, each MPI process introduces an additional row at the top
 and bottom of the grid---these are called {\em ghost} rows and are represented by grey
 horizontal stripped elements in Figure \ref{fig:ghostRegions}. These so-called ghost rows were exchanged using the
 {\tt Sendrecv()} primitive and are supposed to contain vertically stripped rows of their
 left neighbour. For simplicity reasons, only one-sided exchange of rows is shown in Figure \ref{fig:ghostRegions}.
 Listing \ref{list:gameoflifeMPI} depicts a sketch of the parallel implementation of Conway's Game of Life code.

 \begin{lstlisting}[caption=Conway's Game of Life using MPJ Express, label={list:gameoflifeMPI}]
  int cells[][]= new int[B+2][N];
  int sums[][] = new int[B][N];
  while(true) {
    int next=(me + 1)%P;
    int prev=(me - 1 + P)%P;
    MPI.COMM_WORLD.Sendrecv(cells[B], 0, N, MPI.INT,next, 0, cells[0], 0, N, MPI.INT, prev, 0);
    MPI.COMM_WORLD.Sendrecv(cells[1], 0, N, MPI.INT,prev, 0, cells[B+1], 0, N, MPI.INT, next, 0) ;
    // Sum Phase
    for(int i = 1 ; i < B+1 ; i++)
      for(int j = 0 ; j < N ; j++)
        sums[i][j]=sum of all neighbouring cells
    // Update Phase
    for(int i = 0 ; i < B ; i++)
      for(int j = 0 ; j < N ; j++)
        cells[i][j]=update(cells[i][j], sums[i][j]);
  }
\end{lstlisting}

\textbf{Laplace Equation Solver:} Another synchronous computation discussed in the context of
distributed memory parallel programming was the Laplace Equation Solver. The partitioning and
communication patterns exhibited by this application is similar to Conway's Game of Life. Ghost
regions were introduced as an optimization and were communicated using {\tt Sendrecv()} communication
primitive before the actual computational loop executes.

Towards the end, the course also introduced collective communication primitives provided by
MPJ Express---these are depicted in Table \ref{tab:collectiveCommunication}. Students were invited
to re-write distributed memory parallel codes to make use of MPI collective communication instead of
relying on point-to-point communication. Students observed that number of lines of code were
significantly reduced by exploiting specialized collective operations like {\tt Bcast()},
{\tt Reduce()}, {\tt Gather()}, {\tt Allreduce()}, and {\tt Allgather()}. MPJ Express also
provides a global synchronization primitive called {\tt Barrier()}.

\begin{table}[!h]
\centering
\caption{Collective Communication Primitives}
\label{tab:collectiveCommunication}
\begin{tabular}{|m{0.10\textwidth}|m{0.31\textwidth}|} \hline
%\begin{tabular}{|l|p{'5cm'}|} \hline
\multicolumn{1}{|c|}{\cellcolor[gray]{0.8}{\bf Operation}} & \multicolumn{1}{|c|}{\cellcolor[gray]{0.8}{\bf Description}}\\ \hline
{\tt bcast()} (Broadcast)   & One process send message to all other processes\\ \hline
{\tt reduce()}   & Reverse operation of {\tt bcast()} \\ \hline
{\tt Scatter()}  & Distributes distinct messages from one process to all other processes\\ \hline
{\tt Gather()}   & Reverse operation of {\tt Scatter()}\\ \hline
{\tt Allgather()}& Combined operation of {\tt Gather()} and {\tt scatter()}\\ \hline
{\tt Barrier()}  & Creates a barrier synchronization in a group\\ \hline
\end{tabular}
\end{table}

\subsection{Advanced Topics in Parallel Programming}

This sub-section outlines advanced topics including the MapReduce
programming model using Hadoop and the General Purpose Computing on
Graphics Processing Units (GPGPU).

\textbf{Apache Hadoop:} As part of this section, the students were
motivated with the need for MapReduce \cite{MapReduceGoogle} programming model. The central
idea behind this programming model is to process large amounts of
data in a fault-tolerant manner on inexpensive hardware. This is
especially attractive for small-to-medium enterprises and researchers
who cannot afford expensive hardware typically used in HPC environments.
This method of automatic parallelization is getting very popular in the
industry. One obvious advantage of MapReduce over MPI is that the application
developer is not responsible for explicit data placement and communication.
In our course we reviewed the Apache Hadoop \cite{ApacheHadoop} software, which is an open-source
implementation of the MapReduce programming paradigm in the Java language.
We also discussed that the end-user is responsible for writing application
code, which consists of the map and reduce functions. The map
stage typically processes the input data and produces intermediate
key-value pairs. These key-value pairs are then fed to reduce functions,
which combine/filter/sort intermediate data to produce the final
result. When compared with MPI, the application code is remarkably simpler.

\textbf{Word Count Example:} The usage of the MapReduce API was demonstrated
by discussing the classic word count example originally presented in
\cite{MapReduceGoogle}. In this example, the application code is responsible for counting
the frequency of unique words in a collection of documents. The map function
in the example prepares a list of all words and sets the value to $1$. This
data comprising of intermediate key-value pairs is later fed to the reduce function.
As part of reduce phase computation, repeated words in the intermediate data
are summed up to produce the final result, which contains the frequency of
each word present in the input data.

\textbf{GPU Programming:} This part of the course introduced GPUs as a massively-parallel
parallel programming platform. Initial parts of lectures were dedicated to
noting differences between CPUs and GPUs. On one hand CPUs are latency
optimized but are built with complex and power inefficient hardware. On the
other hand, GPUs have simpler hardware and are bandwidth optimized. GPUs
are typically a good choice for applications that involve minimum branching
and have low communication to computation ratio. Various programming APIs including
CUDA \cite{CUDA} and OpenCL were introduced to students. We also introduced JCuda \cite{JCuda}, which consists
of Java bindings for CUDA. Key concepts of GPU programming were introduced to
students through very simple and primitive examples. This part of the course
was not extensively covered. In future offerings of the course, we plan to
provide more coverage by discussing more concrete examples using the JCuda library.

\section{Conclusions}
\label{sec:conclusion}

This paper reviewed a parallel computing course taught to final
year Software Engineering undergraduate students at NUST Pakistan.
A unique feature of the course was that Java was used as the principle
parallel programming language throughout the course. The reason for
preferring Java over traditional languages like C and Fortran was that
it is a modern object-oriented programming language with advanced
features including garbage collection and binary portability.
Another, interesting, argument in favour of Java is that it is
taught as one of the major languages in many Universities around
the globe.

The course was divided into three sections. The first
section---shared memory parallel programming---used Java threads API
to parallelize embarrassingly parallel and synchronous computations
on  multicore and Symmetric Multi-Processor (SMP) systems. The second
section---distributed memory parallel programming---used a Java MPI
library named MPJ Express to parallelize a variety of HPC workloads on
commodity clusters and network of computers. The course did not rely
on any dedicated HPC platform. Instead the students used lab computers
connected using Gigabit Ethernet for executing parallel Java codes.
For this purpose, a custom version of the MPJ Express software was
released\footnote{http://sourceforge.net/p/mpjexpress/mailman/message/ 32311993/}
that was capable of executing on a network of computers with no shared filesystem.
Typically MPI libraries rely on shared storage medium to execute parallel jobs.
The third and the final section covered advanced topics including
the MapReduce programming model using Hadoop and the General Purpose
Computing on Graphics Processing Units (GPGPU).

%To carry out distributed memory programming exercises no dedicated cluster was used instead students were encouraged to use teaching lab PCs to setup a local cluster with non-shared file system. Since students were using MPJ Express which supports running of distributed memory applications on non-shared file system. Earlier versions of MPJ Express (0.41 and before) do not have this support. To make this possible a special version of MPJ Express was developed and released before start of this course which enabled students to run Java MPI applications on non-share file systems. MPJ Express transparently copies the user application executable to each compute node's local storage and runs. This functionality makes it easy and encouraging to run distributed memory parallel applications where NFS is normally not configured or a dedicated cluster is not available.

%
% The following two commands are all you need in the
% initial runs of your .tex file to
% produce the bibliography for the citations in your paper.
%\bibliographystyle{unsrt}
%\bibliography{EduHPC_Submission}  % sigproc.bib is the name of the Bibliography in this case

\begin{thebibliography}{10}

\bibitem{herb}
Herb Sutter and James Larus.
\newblock Software and the concurrency revolution.
\newblock {\em Queue}, 3(7):54--62, September 2005.

\bibitem{threadsJava}
Scott Oaks and Henry Wong.
\newblock {\em Java Threads, Third Edition}.
\newblock O'Reilly Media, Inc., 3 edition, 2004.

\bibitem{openmp}
Robit Chandra, Leonardo Dagum, Dave Kohr, Dror Maydan, Jeff McDonald, and
  Ramesh Menon.
\newblock {\em Parallel Programming in OpenMP}.
\newblock Morgan Kaufmann Publishers Inc., 2001.

\bibitem{cilkplus}
Matteo Frigo, Charles~E. Leiserson, and Keith~H. Randall.
\newblock The implementation of the {C}ilk-5 multithreaded language.
\newblock In {\em Proceedings of the ACM SIGPLAN '98 Conference on Programming
  Language Design and Implementation (PLDI)}, pages 212--223, 1998.

\bibitem{Shafi2009}
Aamir Shafi, Bryan Carpenter, and Mark Baker.
\newblock Nested parallelism for multi-core {HPC} systems using {J}ava.
\newblock {\em Journal of Parallel and Distributed Computing}, 69(6):532 --
  545, 2009.

\bibitem{carpenter}
Bryan Carpenter, Geoffery Fox, Sung-Hoon Ko, and Sang Lim.
\newblock { mpiJava 1.2: API Specification}.
\newblock Technical report, Northeast Parallel Architectures Center, Syracuse
  University, October 1999.
\newblock
  http://www.hpjava.org/reports/mpiJava-spec/mpiJava\-spec/mpiJava-spec.html.

\bibitem{Taboada2012}
Guillermo~L. Taboada, Juan Touri\~{n}o, and Ram\'{o}n Doallo.
\newblock {F-MPJ: Scalable Java Message-passing Communications on Parallel
  Systems}.
\newblock {\em J. Supercomput.}, 60(1):117--140, April 2012.

\bibitem{openmpiJava}
Oscar Vega-Gisbert, Jose~E. Roman, Siegmar Gro\ss, and Jeffrey~M. Squyres.
\newblock Towards the {A}vailability of {J}ava {B}indings in {O}pen{MPI}.
\newblock In {\em Proceedings of the 20th European MPI Users' Group Meeting},
  EuroMPI '13, pages 141--142. ACM, 2013.

\bibitem{ApacheHadoop}
Tom White.
\newblock {\em Hadoop: The Definitive Guide}.
\newblock O'Reilly Media, Inc., 1st edition, 2009.

\bibitem{MapReduceGoogle}
Jeffrey Dean and Sanjay Ghemawat.
\newblock Mapreduce: A {F}lexible {D}ata {P}rocessing {T}ool.
\newblock {\em Commun. ACM}, 53(1):72--77, January 2010.

\bibitem{CUDA}
John Nickolls, Ian Buck, Michael Garland, and Kevin Skadron.
\newblock Scalable {P}arallel {P}rogramming with {CUDA}.
\newblock {\em Queue}, 6:40--53, March 2008.

\bibitem{JCuda}
Yonghong Yan, Max Grossman, and Vivek Sarkar.
\newblock {JCUDA}: A {P}rogrammer-{F}riendly {I}nterface for {A}ccelerating
  {J}ava {P}rograms with {CUDA}.
\newblock In {\em Proceedings of the 15th International Euro-Par Conference on
  Parallel Processing}, Euro-Par '09, pages 887--899. Springer-Verlag, 2009.

\end{thebibliography}
% You must have a proper ".bib" file
%  and remember to run:
% latex bibtex latex latex
% to resolve all references
%
% ACM needs 'a single self-contained file'!
%
%APPENDICES are optional

\balancecolumns
% That's all folks!
\end{document}